\documentclass[11pt,letterpaper]{article}
\pdfoutput=1
\usepackage{jheppub}
\usepackage{bbm}
\usepackage{mathrsfs}
\usepackage{slashed}
\usepackage{caption}
\usepackage{epstopdf}
\usepackage[normalem]{ulem}
\usepackage[bottom]{footmisc}
\usepackage{subcaption}
\usepackage{bbold}
\usepackage{titlesec}
\usepackage{threeparttable}
\usepackage{booktabs}
\usepackage{changepage}
\usepackage[utf8]{inputenc}
\usepackage{dsfont}

\usepackage{grffile}
 
\usepackage{graphicx}  
\usepackage{dcolumn}   
\usepackage{bm}        
\usepackage{amssymb}   
\usepackage{setspace}
\usepackage{amsmath, amssymb, setspace}
\usepackage{array}
\usepackage{booktabs}
\usepackage{caption}
\usepackage{indentfirst}
\usepackage{float}
\usepackage{lmodern}
\usepackage{multirow}
\usepackage{soul}
\usepackage[normalem]{ulem}
\usepackage{xcolor}

\title{\huge{Primordial Black Hole Sterile Neutrinogenesis: Sterile Neutrino Dark Matter Production Independent of Couplings}}

\author[a]{Muping Chen,}
\author[a]{Graciela B. Gelmini,}
\author[b]{Philip Lu,}
\author[c,d,e,f]{and Volodymyr Takhistov} 

\affiliation[a]{Department of Physics and Astronomy, UCLA,
475 Portola Plaza, Los Angeles, CA 90095, USA}
\affiliation[b]{Center for Theoretical Physics, Department of Physics and Astronomy, Seoul National University, 1 Gwanak-ro, Gwanak-gu, Seoul 08826, Korea}
\affiliation[c]{International Center for Quantum-field Measurement Systems for Studies of the Universe and Particles (QUP,WPI), High Energy Accelerator Research Organization (KEK), Oho 1-1, Tsukuba, Ibaraki 305-0801, Japan}
\affiliation[d]{Theory Center, Institute of Particle and Nuclear Studies (IPNS),
High Energy Accelerator Research Organization (KEK), Tsukuba 305-0801, Japan} 
\affiliation[e]{Graduate University for Advanced Studies (SOKENDAI), \\
1-1 Oho, Tsukuba, Ibaraki 305-0801, Japan}
\affiliation[f]{Kavli Institute for the Physics and Mathematics of the Universe (WPI), Chiba 277-8583, Japan}

\emailAdd{mpchen@physics.ucla.edu}
\emailAdd{gelmini@physics.ucla.edu}
\emailAdd{philiplu11@gmail.com}
\emailAdd{vtakhist@post.kek.jp} 

\abstract{ 
Sterile neutrinos ($\nu_s$s) are well-motivated and actively searched for hypothetical neutral particles that would mix with the Standard Model active neutrinos. They are considered prime warm dark matter (DM) candidates, typically when their mass is in the keV range, although they can also be hot or cold DM components. 
We discuss in detail the characteristics and phenomenology of $\nu_s$s that minimally couple only to active neutrinos and are produced in the evaporation of early Universe primordial black holes (PBHs), a process we called ``PBH sterile neutrinogenesis". Contrary to the previously studied $\nu_s$ production mechanisms, this novel mechanism does not depend on the active-sterile mixing. The resulting $\nu_s$s have a distinctive spectrum and are produced with larger energies than in typical scenarios. This characteristic enables $\nu_s$s to be WDM in the unusual $0.3$ MeV to $0.3$ TeV mass range, if PBHs do not matter-dominate the Universe before evaporating. When PBHs matter-dominate before evaporating, the 
possible coincidence of induced gravitational waves associated with PBH evaporation and astrophysical X-ray observations from $\nu_s$ decays constitutes a distinct signature of our scenario.
}

\begin{document}
\preprint{KEK-QUP-2023-0035, KEK-TH-2584, KEK-Cosmo-0334, IPMU23-0049}
 \maketitle
\flushbottom

\section{Introduction}  

Dark matter (DM), which composes about $85\%$ of all the matter in the Universe (see e.g.~\cite{Bertone:2016nfn,Gelmini:2015zpa}), has thus far been detected only through its gravitational interactions and its nature remains mysterious. It is known that the bulk of the DM must be either warm (WDM) or cold (CDM), while a hot (HDM) component can at most be subdominant,  less than 10\% of the total DM. the names HDM, WDM and CDM denote DM particles that are respectively either relativistic, becoming non-relativistic or already non-relativistic when galaxy cores start forming, which happens at a temperature  of about 1 keV. 

The Standard Model (SM) contains active neutrinos characterized by their three flavors $e, \mu$ or $\tau$. These neutrinos couple to the weak gauge bosons and experience the full strength of weak interactions. While within the SM neutrinos are massless, it is known they have non-zero masses that is much smaller than all the other SM particles~\cite{Super-Kamiokande:1998kpq}. Thus, new physics beyond SM is required to explain the observed active neutrino masses. Many models aiming to address the neutrino masses hypothesize the existence of ``sterile" neutrinos ($\nu_s$s), new fermions that would not directly couple to the weak gauge bosons and in the simplest models only couple to the SM through their mixing with active neutrinos. $\nu_s$s are necessary for see-saw mechanisms~\cite{Yanagida:1979as,Gell-Mann:1979vob,Yanagida:1980xy} that in minimal extensions of the SM  generate the observed small neutrino masses. For simplicity, here we shall consider a $\nu_s$ whose only coupling to the SM is a mixing $\sin\theta$ with just one of the active neutrinos $\nu_a$. 

Sterile neutrinos are associated with a rich phenomenology. They constitute a prime WDM candidate, usually if they have $\mathcal{O}$(keV) mass~(see e.g.~\cite{Boyarsky:2018tvu}). Such $\nu_s$s could be efficiently detected through X-ray emission lines coming from  galaxy clusters or the dark halo of our galaxy, due to the two body decay mode $\nu_s \to \nu_a \gamma$ into monochromatic photons. A putative detection of such a line has been claimed at 3.5 keV~\cite{Bulbul:2014sua,Boyarsky:2014jta}  
that could be due to decays of $\nu_s$s with mass $m_s =7.1$~keV, although this interpretation has been challenged~(e.g.~\cite{Jeltema:2014qfa,Dessert:2018qih,Dessert:2023fen}). Assuming that $\nu_s$s constitute all of the DM, the mixing necessary to explain the 3.5 keV line is $\sin^2 2\theta \simeq 7 \times 10^{-11}$~\cite{Bulbul:2014sua,Boyarsky:2014jta} (and would increase inversely to the fraction of the DM consisting of $\nu_s$s). 
Heavier decaying $\nu_s$s have also been proposed to alleviate the discrepancies in recent Hubble constant measurements describing the rate of Universe's expansion~(e.g.~\cite{Gelmini:2019deq,Gelmini:2020ekg,Boyarsky:2021yoh}). Heavy $\nu_s$s are 
sometimes also referred to as ``heavy neutral leptons'' (see e.g.~\cite{Abdullahi:2022jlv} for review).

Sterile neutrinos can be produced in the early Universe through active-$\nu_s$ flavor oscillations and interactions that act as measurements. This is the Dodelson-Widrow (DW)~\cite{Dodelson:1993je,Barbieri:1989ti,Barbieri:1990vx} mechanism.  
The resulting energy-momentum spectrum produced is expected to be close to thermal and the mechanism is already strongly constrained by observations, in particular X-ray data~(e.g.~\cite{Palazzo:2007gz,Horiuchi:2013noa,Perez:2016tcq,Dessert:2018qih,Roach:2022lgo}). In the presence of a very large lepton asymmetry, which we do not assume here, active-sterile oscillations are resonant and the produced $\nu_s$s have a colder spectrum~\cite{Shi:1998km}. This fares better with respect to X-ray limits. A variety of other mechanisms, such as inflaton decays~\cite{Shaposhnikov:2006xi}, sterile neutrino self-interactions~\cite{Bringmann:2022aim}, freeze-in in extended particle models~\cite{Asaka:2006ek,Roland:2014vba}, and extended gauge symmetries~\cite{Dror:2020jzy,Bezrukov:2009th}, assume that $\nu_s$s have additional interactions, which we do not assume to exist in the present paper. All these different mechanisms result in the production of distinctive cosmological $\nu_s$ abundances and energy-momentum spectra. The predictions are also sensitive to the details of cosmological scenarios (see e.g.~\cite{Gelmini:2019esj,Gelmini:2019wfp,Gelmini:2019clw,Chichiri:2021wvw}).

Recently in Ref.~\cite{Chen:2023lnj}, we presented the main characteristics of $\nu_s$s as DM constituents when they are predominantly produced in
the evaporation of early Universe primordial black holes (PBHs), for $\nu_s$s that couple to the SM only through their mixing with a $\nu_a$.
This $\nu_s$ production
mechanism, which we called \textit{PBH sterile neutrinogenesis}, had not been previously considered for $\nu_s$s as DM components. Black hole evaporation has been recently identified as efficient production source of other DM and dark sector particles as well, all mostly or entirely decoupled from the SM, and for production of dark radiation (e.g.~\cite{Matsas:1998zm,Bell:1998jk,Khlopov:2004tn,Fujita:2014hha,Lennon:2017tqq,Morrison:2018xla,Hooper:2019gtx,Baldes:2020nuv,Masina:2020xhk,Gondolo:2020uqv,Sandick:2021gew,Cheek:2021odj,Arbey:2021ysg,Marfatia:2022jiz,Kim:2023ixo}).

In this work, expanding 
Ref.~\cite{Chen:2023lnj}, we describe in detail all  aspects of PBH sterile neutrinogenesis as well as its observational consequences.
There are several essential differences between the characteristics of
 $\nu_s$s produced in PBH sterile neutrinogenesis and those produced in other ways. To start with, PBH evaporation is a purely gravitational process and thus $\nu_s$ production in this model is independent  of  the active-sterile mixing. Hence, even $\nu_s$s with a very small mixing can compose a non-negligible fraction of the DM. Also,  PBH evaporation  produces $\nu_s$s  with a distinctive 
spectrum, very different from the close to thermal spectrum that is typical in other cosmological production mechanisms. During evaporation particles are produced at the energy scale of the black hole Hawking temperature, which is very high for all PBHs that evaporate in the early Universe. Hence, $\nu_s$s have initially higher energies and constitute HDM, WDM or CDM components for much higher  mass $m_s$ values than characteristic of other production mechanisms. Furthermore, PBH sterile neutrinogenesis may lead to intriguing novel observational signatures. In the case in which evaporation occurs after PBH matter-domination, PBH sterile neutrinogenesis enables a coincidence of X-ray signals  originating from astrophysical sources such as galaxy clusters,  and gravitational waves (GW) which could be detected in future observatories or cosmological probes. These coincident signals would distinguish this mechanism from others.  

The present work is organized as follows.
In Sec.~\ref{sec:PBH} we give an overview of PBH formation and evaporation in the early Universe. In Sec.~\ref{sec:sterilenu} we discuss the relic abundance, energy spectrum and non-thermalization condition of $\nu_s$s due to PBH sterile neutrinogenesis. We discuss constraints and observational signatures for the case when PBHs evaporate in the initial radiation-dominated era in Sec.~\ref{sec:rdom} and for the case in which PBH matter-dominate before evaporating in Sec.~\ref{sec:mdom}. We conclude in Sec.~\ref{sec:conclusions}. We use natural units $c = \hbar = 1$ throughout.

\section{Evaporating black holes in early Universe}
\label{sec:PBH}

We consider a broad range of initial PBH masses, their evaporation temperatures and their abundances at evaporation. These serve as initial conditions and inputs for sterile neutrino DM production in PBH sterile neutrinogenesis. 
Hence, we consider the existence of a population of PBHs in the Early Universe with the appropriate characteristics that must have been produced early enough. Many PBH production scenarios have been proposed (e.g.~\cite{Zeldovich:1967,Hawking:1971ei,Carr:1974nx,Carr:1975qj,GarciaBellido:1996qt,Green:2000he,Frampton:2010sw,Green:2016xgy,Sasaki:2018dmp,Cotner:2018vug,Cotner:2019ykd,Kusenko:2020pcg,Carr:2020gox,Green:2020jor,Escriva:2022duf,Lu:2022yuc}), enabling PBH sterile neutrinogenesis to be realized within a multitude of theories.
Here, we briefly expand on the most frequently studied PBH formation scenario due to direct gravitational collapse upon Hubble horizon crossing of
highly overdense regions seeded by inflationary inhomogeneities in the primordial Universe (see e.g.~\cite{Sasaki:2018dmp,Carr:2020gox,Green:2020jor,Escriva:2022duf} for a review and references therein). 

\subsection{Primordial black hole formation}
\label{PBH-formation}

PBHs can be formed from the
collapse of large density fluctuations 
sourced by inflationary perturbations, when
the curvature power spectrum has a significant enhancement at scales smaller than those associated with cosmic microwave background (CMB) (see e.g.~\cite{Sasaki:2018dmp,Carr:2020gox,Green:2020jor,Escriva:2022duf} for review). We consider this PBH formation mechanism as our reference scenario. The mass of the resulting PBH is a sizable fraction of the Hubble horizon mass $M_H$ at formation during the radiation-dominated era, when the temperature is $T_{\rm form}$~(e.g.~\cite{Sasaki:2018dmp,Carr:2020gox}),
\begin{equation}
\label{eq:MPBH}
    M_{\rm PBH} = \gamma M_H \simeq 1.89\times 10^{7} {\rm g} \left(\frac{\gamma}{0.2}\right)\left(\frac{10^{12}\textrm{ GeV}}{T_{\rm form}}\right)^{2}\left(\frac{106.75}{g_*(T_{\rm form})}\right)^{1/2}~,
\end{equation}
where $g_*(T)$ is the number of relativistic degrees of freedom of the radiation bath at temperature $T$ (106.75 when all the SM particles are in equilibrium), and $\gamma$ is the fraction of the mass within the horizon that falls into the PBH. The precise value of $\gamma$ depends on the details of the collapse, with simple analytic calculations suggesting $\gamma \simeq 0.2$~\cite{Carr:1975qj} that we take as its characteristic value.

In order for PBHs to form, the fluctuations must exceed the critical (threshold) overdensity value $\delta_c$, which sensitively depends on initial conditions and assumptions and is typically $\mathcal{O}(0.5)$~(see~e.g.~\cite{Musco:2018rwt,Sasaki:2018dmp} for discussion). The fraction of the total energy density in PBHs at formation,  assuming radiation-domination, is
\begin{equation}
\label{eq:beta}
    \beta \simeq \frac{\rho_{\rm PBH}(T_{\rm form})}{\rho_{\rm Rad}(T_{\rm form})} = \gamma \int_{\delta_c}^{\infty} P(\delta) d\delta~,
\end{equation}
where $\rho_{\rm Rad}(T)=\pi^2g_*(T)T^4/30$ is the radiation energy density at temperature $T$.
Here $P(\delta)$ is the probability distribution of primordial density fluctuations. To keep the discussion general, we do not restrict ourselves to a specific model of inflation and take  $\beta$ as a phenomenological parameter. 

Different formation mechanisms can lead to a variety of resulting PBH features, including distributions in PBH masses and spins. For simplicity, we assume a monochromatic PBH mass function throughout. Since the PBH evaporation time and temperature depend sensitively on the PBH mass, as discussed below in Sec.~\ref{sec:evaporation}, the PBH mass distribution function can impact $\nu_s$ production as well as other associated processes, like the generation of induced GWs~(e.g.~\cite{Domenech:2021wkk}). Further, here we only consider non-spinning Schwarzschild PBHs, although in different models PBHs can form with a sizable spin~(e.g.~\cite{Harada:2017fjm,Cotner:2016cvr,Cotner:2018vug,Cotner:2019ykd}), which affects the PBH evaporation.
We leave detailed analysis of these scenarios for future work.

\subsection{Primordial black hole evaporation}
\label{sec:evaporation}
 
PBHs emit Hawking radiation with an approximately blackbody spectrum at the Hawking temperature~\cite{Hawking:1975vcx,Page:1976df,Page:1976ki,Page:1977um,MacGibbon:1990zk}
\begin{equation}
\label{eq:tbh}
    T_{\rm PBH} =
    \frac{1}{8 \pi M_{\rm PBH}}=1.06\times10^5 \textrm{ GeV} \left(\frac{10^8 {\rm g}}{M_{\rm PBH}}\right)~. 
\end{equation}
Evaporating black holes emit all particles in the spectrum whose mass is smaller than 
the Hawking temperature $T_{\rm PBH}$, regardless of any other interactions besides gravity.
The emission of low energy states is suppressed, especially for higher spin particles, but the spectrum approaches that of a blackbody 
for high energies~\cite{MacGibbon:1990zk}. The emission rate is given by~\cite{MacGibbon:1990zk,MacGibbon:1991tj}
\begin{equation}
\label{eq:emrate}
    \frac{dM_{\rm PBH}}{dt} = -7.6\times 10^{8}~ \frac{\rm g}{\rm s}~ g_{H}(T_{\rm BH})\left( \frac{10^8{\rm  g}}{M_{\rm PBH}}\right)^{2}~,
\end{equation}
where the factor $g_H$  counts the number of species with mass below the Hawking temperature weighted  by their relative emission rates (normalized to spin-1/2 particles, as e.g. in Ref.~\cite{Hooper:2019gtx})
\begin{equation}
\label{eq:gh}
 g_H = \sum_i w_i g_{H,i}~,
~~~g_{H,i}=\left\{
\begin{aligned}
    &=  1.82,~s=0 \\
    &=  1.0,~~s=1/2 \\
    &=  0.41,~s=1 \\
    &=  0.05,~s=2
\end{aligned}
\right.
\end{equation}
with $w_i$ being the number of internal degrees of freedom. Including only the particles of the SM, the value of $g_H$   is 108, and 
the contribution to $g_H$  due to $\nu_s$s is $g_S= 2$. Thus, the total value of the emission factor in our model is $g_H=110$.

The evaporation rate~\cite{MacGibbon:1990zk} is such that the  PBH lifetime is 
\begin{equation}
\label{eq:lifetime}
    \tau_{\rm evap} \simeq 3.9\times10^{-4}~{\rm s} \left(\frac{M_{\rm PBH}}{10^8{\rm g}}\right)^3 \left(\frac{110}{g_H}\right)~.
\end{equation}
The evaporation rate increases rapidly as the PBH evaporates, with most of the emission happening at the end of the PBH lifetime. Thus,
for simplicity, in the following we make the common assumption of instantaneous reheating, in which the redshift effects on particles emitted during the evaporation process is negligible. In this case, the average energy of the emitted $\nu_s$s is $\langle p \rangle \simeq 6.3 ~ T_{\rm PBH}$~\cite{Baldes:2020nuv}.

We assume that PBHs are formed in a radiation dominated Universe with initial energy density fraction $\beta < 1$.  If PBHs evaporate before they come to dominate the energy density of the Universe, 
the expansion rate is $H^2=8\pi \rho_{\rm Rad}/(3 M_{\rm Pl}^2) =1/(4t^2)$. Since $H= 1/2t$,
at the time of evaporation $t=\tau_{\rm evap}$ the temperature is
\begin{equation}
\label{eq:trhevap}
\begin{aligned}
        T_{\rm evap} & \simeq 43 \textrm{ MeV}  \left(\frac{10^8 {\rm g}}{M_{\rm PBH}}\right)^{3/2}\left(\frac{10.75}{g_*(T_{\rm evap})}\right)^{1/4}\left(\frac{g_H}{110}\right)^{1/2}\\
       & \simeq 39~{\rm MeV}\left(  \frac{T_{\rm PBH}}{\rm 10^5 GeV} \right)^{3/2}\left( \frac{10.75}{g_*(T_{\rm evap})} \right)^{1/4}\left(\frac{g_H}{110}\right)^{1/2}~.
\end{aligned}
\end{equation}
Between PBH formation and evaporation,  the PBH energy density increases relative to the radiation density, since PBHs are non-relativistic massive objects, and PBHs could matter-dominate the Universe before evaporating. 

If PBHs never dominate the energy density of the Universe and evaporate during the radiation dominated era, a case we call ``RD",  the PBH fraction of the total energy density (assuming $\rho_{\rm Total} \simeq \rho_{\rm Rad}$)  at evaporation is
\begin{align}
\begin{split}
\label{eq:fevapfrac}
    f_{\rm evap} = &~ \frac{\rho_{\rm PBH}(T_{\rm evap})}{\rho_{\rm Total}(T_{\rm evap})}\simeq \frac{\rho_{\rm PBH}(T_{\rm evap})}{\rho_{\rm Rad}(T_{\rm evap})}  \simeq \beta \left(\frac{T_{\rm form}}{T_{\rm evap}}\right)\left(\frac{g_*(T_{\rm form})}{g_*(T_{\rm evap})}\right)^{1/3} \\ \simeq &~ 0.2 \left(\frac{\beta}{10^{-12}}\right)\left(\frac{T_{\rm form}}{10^{10} \textrm{ GeV}}\right)\left(\frac{M_{\rm PBH}}{10^8 {\rm g}}\right)^{3/2} \left(\frac{g_*(T_{\rm form})}{g_*(T_{\rm evap})}\right)^{1/3}\left(\frac{g_*(T_{\rm evap})}{10.75}\right)^{1/4}~,
    \end{split}
\end{align}
where $T_{\rm form}$ is the temperature at PBH formation during radiation-dominated era.
Considering PBH formation scenario via collapse of density fluctuations as described in Sec.~\ref{PBH-formation}, $T_{\rm form}$ can be written in terms of $M_{\rm PBH}$  using Eq.~\eqref{eq:MPBH}, resulting in
\begin{equation}
\label{eq:fevaprad}     
    f_{\rm evap} \simeq 0.46 \left(\frac{\beta}{10^{-13}}\right) \left(\frac{M_{\rm PBH}}{10^8 {\rm g}}\right) \left(\frac{g_*(T_{\rm evap})}{10.75}\right)^{-1/12}~.
\end{equation}

The condition for the RD case is that PBHs are subdominant at evaporation, i.e.  $f_{\rm evap}< 1$. Using  Eq.~\eqref{eq:fevapfrac} this condition is 
\begin{equation}
\label{eq:domcond1}
    T_{\rm evap} > \beta ~T_{\rm form} \left(\frac{g_*(T_{\rm form})}{g_*(T_{\rm evap})}\right)^{1/3}~.
\end{equation}
In the PBH formation scenario of Sec.~\eqref{PBH-formation}, using Eq.~\eqref{eq:fevaprad},  Eq.~\eqref{eq:domcond1} becomes
\begin{equation}
\label{eq:domcond}
    \beta < 3.8 \times10^{-14} \left(\frac{10^8 {\rm g}}{M_{\rm BH}}\right) \left(\frac{g_*(T_{\rm evap})}{g_*(T_{\rm form})}\right)^{1/3}~.
\end{equation}

PBHs can matter-dominate the Universe before evaporating, a case we call ``PD". This occurs if the condition in  Eq.~\eqref{eq:domcond1}, or Eq.~\eqref{eq:domcond}, is not fulfilled.
In the PD case, the PBH energy density at evaporation $\rho_{\rm PBH-evap}$ is approximately the total density (i.e. $f_{\rm evap} \simeq 1$) defined through the expansion rate when $t=\tau_{\rm evap}$ in a matter-dominated epoch (assuming this epoch is long enough to neglect the earlier RD period),  i.e.   $H = \sqrt{8\pi \rho_{\rm PBH-evap}/(3 M_{\rm Pl}^2)} =2/(3 \tau_{\rm evap})$.
Due to the very fast thermalization of all SM particles, at the moment of evaporation nearly the entire PBH energy density
becomes instantaneously a radiation bath with temperature $T_{\rm RH}$ defined by $\rho_{\rm PBH-evap}= \rho_{\rm Rad} (T_{\rm RH})$, with
\begin{equation}
\label{eq:trhm}
\begin{aligned}
        T_{\rm RH}& \simeq 55 \textrm{ MeV} \left(\frac{10^{8}~ {\rm g}}{M_{\rm PBH}}\right)^{3/2} \left(\frac{10.75}{g_{*}(T_{\rm RH})}\right)^{1/4}\left(\frac{g_H}{110}\right)^{1/2}~\\
         & \simeq 50~{\rm MeV}\left(  \frac{T_{\rm PBH}}{\rm 10^5~GeV} \right)^{3/2}\left( \frac{10.75}{g_*(T_{\rm RH})} \right)^{1/4}\left(\frac{g_H}{110}\right)^{1/2}~.
\end{aligned}
\end{equation}

We impose two conditions on the  evaporation or reheating  temperature, in the RD case and PD case respectively. The first is that it cannot be lower than $\sim5$~MeV due to limits imposed by Big Bang Nucleosynthesis (BBN)~\cite{Kawasaki:1999na,Kawasaki:2000en,Hannestad:2004px,Ichikawa:2005vw,Ichikawa:2006vm,DeBernardis:2008zz,deSalas:2015glj,Hasegawa:2019jsa}.
The second condition is that, as a consistency requirement of our model, this temperature cannot be as large as the black hole temperature at the moment of evaporation, $T_{\rm PBH}$, thus we chose it to be at most 0.1~$T_{\rm PBH}$.
This second condition implies $T_{\rm evap} <1.94\times 10^{15}$~GeV in the RD case,  $T_{\rm RH} <1.44\times 10^{15}$GeV in the PD case. Applying these two conditions, the PBH mass is restricted to be in the range $5.41\times 10^{-4}$g $<M_{\rm PBH}<4.23\times 10^{8}$g in the RD case, and  $7.29\times 10^{-4}$g$<M_{\rm PBH} <4.67\times 10^{8}$g in the PD case.  This is the range of the vertical axis in Figs.~\ref{fig:Raddom} and \ref{fig:PBHdom} presented below. 
The region of $M_{\rm PBH} < 0.1$g in these figures (above the dashed brown line) is restricted by the upper bound on the energy scale of inflation $\rho_{\rm inf} < (1.6 \times 10^{16} \textrm{GeV})^4$ due to constraints on the tensor-to-scalar ratio~\cite{Planck:2018jri}, and consequent upper limit on $T_{\rm evap}$ or $T_{\rm RH}$. However, we note that light PBHs could be created at smaller temperatures in different PBH production mechanisms, see e.g.~\cite{Cotner:2016cvr, Deng:2017uwc,Cotner:2018vug,Cotner:2019ykd,Kawana:2021tde}, thus the upper limit
on temperature would not reject them. 

\section{Sterile neutrino  production}
\label{sec:sterilenu}
Particles are produced through PBH evaporation due to their gravitational interaction 
once the Hawking temperature exceeds the particle mass. Since we require  $M_{\rm PBH} \lesssim 4\times 10^8$ g (see the end of 
Sec.~\ref{sec:evaporation}), which  through Eq.~\eqref{eq:tbh} corresponds to Hawking temperatures $T_{\rm PBH}\gtrsim 3\times10^4$ GeV, 
$\nu_s$s with mass $m_s <10^4$ GeV are 
 ultra-relativistic at emission.
Although the emission rate has a dependence on fundamental black hole parameters, it is independent of any couplings the particles could have. In contrast to the DW and other $\nu_s$ production mechanisms, i.e due to decays~\cite{Enqvist:1990ad,Enqvist:1990ek,Dodelson:1993je,Shi:1998km}, PBH sterile neutrinogenesis   can produce a relevant relic abundance
even for arbitrarily small active-sterile
mixings. 

We are interested in  $\nu_s$'s that do not significantly interact after being produced in the evaporation of PBHs, and thus their energy just redshifts to the present day. Assuming first that this is so, in this section we present their relic momentum distribution and abundance, and find how small the active-sterile mixing angle must be for $\nu_s$s to be primarily produced through PBH evaporation as opposed to being produced through the DW mechanism.
Then, we find how sizable the active-sterile mixing must be for $\nu_s$s to thermalize. We show that for the DW population to be subdominant the mixing must almost always be below the thermalization range. 

\subsection{Relic momentum distribution}
\label{ssec:spectrum}

\begin{figure}
    \centering
\includegraphics[width=0.7\textwidth]{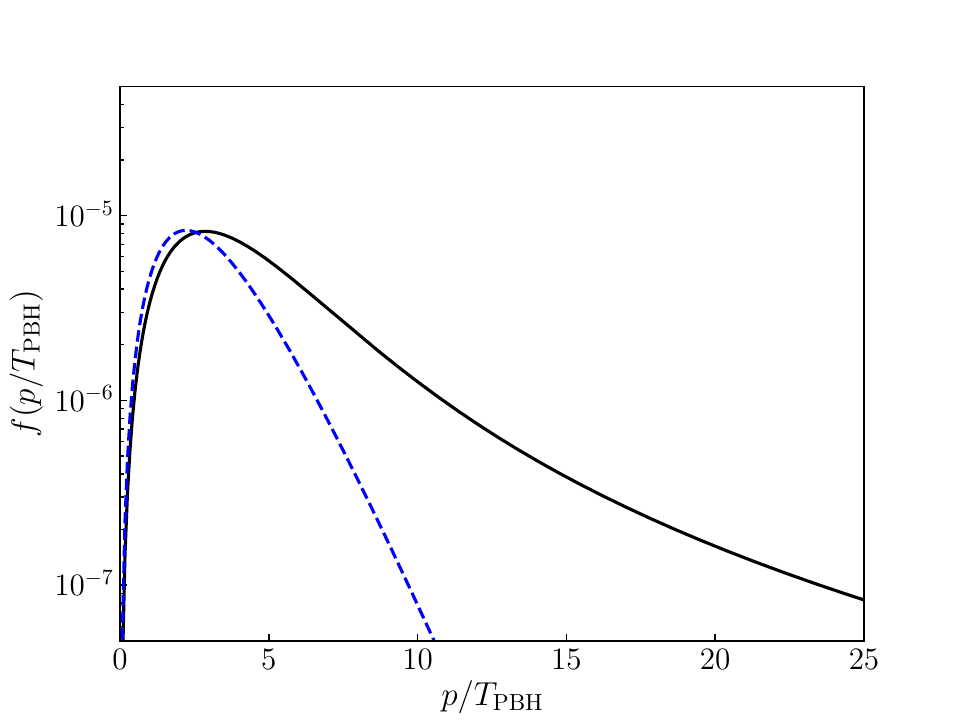}
    \caption{Momentum distribution of $\nu_s$s produced in PBH sterile neutrinogenesis displayed as a function of momentum $p/T_{\rm PBH}$, derived using the instantaneous evaporation approximation~\cite{Baldes:2020nuv} (solid black line). For reference, a Fermi-Dirac spectrum with the same temperature $T_{\rm PBH}$ scaled so that its maximum is at about the same level is also shown  (dashed blue line). }  
    \label{fig:spectrum}
\end{figure}

The primary emission of  PBHs has 
an approximate blackbody spectrum 
with suppression in the low energy-momentum states (see e.g. Fig. 1 of Ref.~\cite{MacGibbon:1990zk}).
For spin $1/2$ particles, the instantaneous  emission spectrum peaks at $E=4.5~ T_{\rm PBH}$.
The time integrated  spectrum assuming an instantaneous reheating (i.e. short enough that redshifting is negligible) can be approximated by two power laws, with $E dN/dE \propto E^3$ for $E\ll T_{\rm PBH}$ and $E dN/dE \propto E^{-2}$ for $E \gg T_{\rm PBH}$, where $N$ is the number of $\nu_s$s emitted (see for example  Refs.~\cite{MacGibbon:1991tj,Baldes:2020nuv}).  This spectrum~\cite{Baldes:2020nuv} (black solid line)  is plotted in Fig.~\ref{fig:spectrum} together with
a Fermi-Dirac distribution (dashed blue line) of the same temperature, scaled (by a factor of order $10^{-3}$) so that the maxima of both spectra are at a similar level for comparison. The spectrum has a long tail, due to the steady rise of the PBH temperature during evaporation, thus the average momentum is larger than the peak.  The average momentum of the emitted $\nu_s$s is $\langle p \rangle \simeq 6.3~T_{\rm PBH}$~\cite{Baldes:2020nuv}.  
The corresponding average scaled dimensionless momentum $\langle \epsilon\rangle$ at evaporation is
\begin{equation}
\label{eq:average-epsilon}
    \langle \epsilon \rangle = \frac{\langle p\rangle}{T_{\rm evap}}\simeq  \frac{6.3~T_{\rm PBH}}{T_{\rm evap}} \simeq 1.5\times10^7 \left(\frac{M_{\rm PBH}}{10^8 {\rm g}}\right)^{1/2}\left( \frac{g_*(T_{\rm evap})}{10.75}\right)^{1/4}~,
\end{equation}
which is much larger than $\langle \epsilon \rangle_{\rm FD} = 3.15$ expected from a Fermi-Dirac spectrum. Eq.~\eqref{eq:average-epsilon} is valid for the RD case, and also for the PD by replacing $T_{\rm evap}$ with $T_{\rm RH}$.  

\subsection{Relic abundance}
\label{ssec:abundance}

 In the RD case, at PBH evaporation the fraction of the total energy density emitted in $\nu_s$s 
is the fraction of the total energy density at evaporation consisting of PBHs, $f_{\rm evap}$ (see Eq.~\eqref{eq:fevapfrac} or Eq.~\eqref{eq:fevaprad}) times the ratio of $\nu_s$ and total degrees of freedom in the Hawking radiation, namely 
\begin{equation}
    \frac{\rho_s(T_{\rm evap})}{\rho_{\rm Rad}(T_{\rm evap})}
    = f_{\rm evap}~\left(\frac{g_S}{g_H}\right) = \frac{f_{\rm evap}}{55}\left(\frac{g_S}{2}\right)\left(\frac{110}{g_H}\right)~.
\end{equation}
In the PD case the same equation holds at reheating, by replacing $f_{\rm evap}=1$.    Recall that $g_S = 2$  for Majorana neutrinos, representing two fermionic degrees of freedom in Eq.~\eqref{eq:gh}. 
In the following, we consider $g_s=2$ and $g_H=110$ and avoid writing $g_s$ and $g_H$ in the equations. 
The number density of $\nu_s$s at evaporation $n_s(T_{\rm evap})= \rho_s(T_{\rm evap})/\langle p\rangle$ is
\begin{equation}
\label{eq:nsevap}
    n_s(T_{\rm evap}) = \frac{f_{\rm evap}}{55}
    ~ \frac{\pi^2 ~g_*(T_{\rm evap}) ~T_{\rm evap}^4}{30 ~ (6.3~ T_{\rm PBH})}~,
\end{equation}
where  as we mentioned in Sec.~\ref{ssec:spectrum}, $\langle p\rangle = 6.3~ T_{\rm PBH}$~\cite{Baldes:2020nuv}. The number density at present ($T_0=2.3\times 10^{-4}$~eV) is
\begin{equation}
\label{eq:ns0}
    n_s(T_0) = n_s(T_{\rm evap})\dfrac{g_0 T_0^3}{g_*T_{\rm evap}^3} = \frac{f_{\rm evap}}{55}
    ~ \frac{\pi^2 ~g_0 T_0^3~T_{\rm evap}}{30 ~ (6.3~ T_{\rm PBH})}~,
\end{equation}
where $g_0 = 3.9$.

The SM particles have gauge couplings and thus rapidly
thermalize into a radiation bath after being emitted at black hole evaporation. However, except for very large mixing angles that we specify in Sec.~\ref{ssec:therm}, $\nu_s$s do not thermalize and their energy just redshifts as the Universe expands. Thus, PBH sterile neutrinogenesis produces a characteristic sparse population of  highly energetic $\nu_s$ with a
hot spectrum that distinguishes it
from other production mechanisms.
 
In the RD case, the present DM fraction consisting of $\nu_s$s due to PBH evaporation is
\begin{equation}
\label{eq:fnr-RD}
\begin{aligned}
    f_s = &~  \frac{\rho_{s}(T_0)}{\rho_{\rm DM}} \simeq 
    2 \times10^{-6}  f_{\rm evap} \left(\frac{m_s}{\rm  keV}\right)\left(\frac{10^8{\rm g}}{M_{\rm PBH}}\right)^{1/2} \left(\frac{10.75}{g_*(T_{\rm evap})}\right)^{1/4} ~\\
    \simeq &~   1 \times 10^{-6}f_{\rm evap}\left( \dfrac{m_s}{\rm keV}\right) \left( \dfrac{T_{\rm evap}}{\rm 5~MeV}\right) ^{{1}/{3}} \left( \dfrac{10.75}{g_{\ast }\left( T_{\rm evap}\right) }\right) ^{{1}/{6}}.
\end{aligned}
\end{equation}
Recall that in the RD case $f_{\rm evap} <1$, which suppresses the $\nu_s$ abundance. In the PD case, the same equation is valid taking $f_{\rm evap}=1$, and  replacing $T_{\rm evap}$ by $T_{\rm RH}$.

Here we impose that the active-sterile neutrino mixing is small enough that $\nu_s$ produced though flavor oscillations are a subdominant population. We will demonstrate that even in this situation, the decays of $\nu_s$s can lead to interesting observable consequences. 

\subsection{Subdominant population produced from oscillations}
\label{ssec:subdominant-DW}

In addition to PBH sterile neutrinogenesis, $\nu_s$s can also be produced through DW mechanism from active-sterile neutrino oscillations (see Appendix~\ref{app:osc} for an overview). This leads to a DM fraction
 $f_{s,osc}$, as given in Eq.~\eqref{eq:fsdmosc}.  Since Hawking radiation emits $\nu_s$s at the same rate for given $m_s$ and $M_{\rm PBH}$ values, independently of their mixing, it can 
constitute the dominant production if the mixing is sufficiently small.
The condition $f_s>f_{s,osc}$ translates into condition on mixing
\begin{equation}
\label{eq:sinRstd}     
    \sin^2(2\theta) < 4 \times10^{-13} 
    f_{\rm evap}\left(\frac{\rm keV}{m_s} \right)\left(\frac{10.75}{g_*(T_{\rm evap})}\right)^{1/6}\left( \frac{g_*(T_{\rm max})}{30}\right)^{3/2}\left( \frac{T_{\rm evap}}{\rm 5~ MeV}\right)^{1/3},
\end{equation}
in the RD case.

The DW production rate has a sharp maximum at a temperature $T_{\rm max}$, given in Eq.~\eqref{eq:tmax}.
Thus, in the PD case, the DW production depends on the value of the reheating temperature. If $T_{\rm RH}> T_{\rm max}$ the  DW production is unsuppressed. With the produced DW DM fraction being $f_{s,osc}$ the condition for PBH neutrinogenesis dominance, $f_s> f_{s,osc}$,   is given by  the same
Eq.~\eqref{eq:sinRstd} with $f_{\rm evap}= 1$ and $T_{\rm evap}$ replaced by $T_{\rm RH}$.

If instead in the PD case $T_{\rm RH} < T_{\rm max}$ (region hatched in cyan in Fig.~\ref{fig:PBHdom}),  this is a version of a  Low Reheating Temperature (LRT) scenario (see e.g. Refs.~~\cite{Gelmini:2004ah,Gelmini:2019esj,Gelmini:2019wfp}) in which the DW production rate is suppressed and the fraction of $\nu_s$s in the DM $f^{\rm LRT}_{osc}$ in Eq.~\eqref{eq:fsLRTosc} is smaller and depends on $T_{\rm RH}$. The condition for PBH neutrinogenesis dominance $f_s >f^{\rm LRT}_{osc}$ in this case becomes
\begin{equation}
\label{eq:sinMLRT}  
\sin^2(2\theta)< 1 \times 10^{-9} f_{\rm evap}
\left(\frac{10.75}{g_*(T_{\rm RH})}\right)^{1/6}\left( \frac{\rm 5~MeV}{T_{\rm RH}}\right)^{8/3}~.
\end{equation}

Since PBH sterile neutrinogenesis results in a high average momentum, $\langle p \rangle= \langle \epsilon \rangle T_{\rm evap}= 6.3 T_{\rm PBH}$~\cite{Baldes:2020nuv}, sterile neutrinos heavier than the active neutrinos $m_s>m_a$, may be relativistic in the present day and constitute dark radiation (DR), i.e. if $m_s< \langle \epsilon \rangle T_0 (3.36/g_*(T_{\rm evap}))^{1/3}$, where $T_0 = 2.3\times10^{-4}\textrm{ eV}$ is the present radiation temperature.  In this case, the present-day energy density of these sterile neutrinos is affected, so that the condition $f_{s,osc} < f_{s}$  for sterile neutrinogenesis to be dominant becomes
\begin{equation}
    \frac{\sin^2 2\theta}{2.9\times10^{-7}} < f_{\rm evap}\left[\frac{\textrm{eV}}{m_s}\right]^2\left[\frac{10.75}{g_*(T_{\rm evap})}\right]^{\frac{1}{3}}\left[\frac{g_*(T_{\rm max})}{10.75}\right]^{\frac{3}{2}}.
\end{equation}

On the other hand, for sterile neutrinos lighter than active neutrinos, $m_s< m_a$, there can be resonant production of sterile neutrinos due to the negative mass difference $\delta m^2 = m_a^2 - m_s^2$, generating a significant lepton asymmetry in the active neutrino just before BBN~\cite{Dolgov:2002wy}. We use the approximation in the small mixing angle regime pertinent to our study, where the generated asymmetry given by (see below Eq. (383) in Ref.~\cite{Dolgov:2002wy})
\begin{equation}
    L_a\simeq 30 \left[\frac{\delta m^2}{2.5\times10^{-3}\textrm{ eV}^2}\right]\left[\frac{\sin^2 2\theta}{10^{-8}}\right]^4.
\end{equation}
The lepton asymmetry is bounded by BBN observations, $L_e < 2\times10^{-3}, L_{\mu,\tau}<0.1$~\cite{Barenboim:2016shh}. Considering the most stringent case of sterile-electron neutrino mixing, the bound for light sterile neutrinos is $\sin^2 2\theta \lesssim 9\times10^{-10} (\delta m^2/2.5\times10^{-3}\textrm{ eV}^2)^{-1/4}$.

The generated lepton asymmetry produced during this resonance phase can be translated into the number density of sterile neutrinos using $n_{\nu_s} \simeq L_{a}n_\gamma$. In contrast to the extremely hot spectrum of sterile neutrinos produced in PBH sterile neutrinogenesis, the resonantly produced sterile neutrinos produced have a cold spectrum with low average momentum, which we estimate to be on the order of $\langle \epsilon \rangle \simeq 1$. If the resonantly-produced sterile neutrinos are non-relativistic at present, $T_{\nu}< m_s$, the condition for PBH sterile neutrinogenesis to be dominant over resonant production is
\begin{equation}
    \frac{\sin^2 2\theta}{6.4\times10^{-10}} <  f_{\rm evap}^{\frac{1}{4}} \left[\frac{0.01\textrm{ eV}}{m_s}\right]^{\frac{1}{4}} \left[\frac{2.5\times10^{-3}\textrm{ eV}^2}{\delta m^2}\right]^{\frac{1}{4}},
\end{equation}
and if the resonantly-produced sterile neutrinos are relativistic at present $T_{\nu}>m_s$, the condition becomes
\begin{equation}
    \frac{\sin^2 2\theta}{1.8\times10^{-9}} <  f_{\rm evap}^{\frac{1}{4}} \left[\frac{2.5\times10^{-3}\textrm{ eV}^2}{\delta m^2}\right]^{\frac{1}{4}}.
\end{equation}

\subsection{Thermalization of sterile neutrinos}
\label{ssec:therm}

Let us first consider the DW process of $\nu_s$ production from a population of active $\nu_a$s, following Eq.~\eqref{eq:Boltzmann}. For very large mixing angles,
the momentum distribution of $\nu_s$s produced via the DW mechanism approaches asymptotically
the $\nu_a$ thermal Fermi-Dirac distribution, as blocking factors saturate, inverse processes in Eq.~\eqref{eq:Boltzmann} become equal and $\nu_s$s are in thermal equilibrium with the $\nu_a$s.
This happens when~\cite{Gelmini:2020duq}
\begin{equation}
\label{eq:thermalization}
    \sin^2 2\theta \geq 
    5 \times10^{-6} \left(\frac{\textrm{keV}}{m_s}\right)\left(\frac{g_* (T_{\rm max})}{10.75}\right)^{1/2}~.
\end{equation}
In Eq.~\eqref{eq:Boltzmann} we have used the
interaction rate and in medium potential valid for temperatures below 100 GeV, for which the weak vector bosons are massive. This is justified since for the neutrino masses we consider, the sharp maximum of the DW production rate happens at temperatures $T_{\rm max}$ that fulfill this condition. In fact Eq.~\eqref{eq:tmax}  shows that  $T_{\rm max} < 100$~GeV for $m_s < 10^3$~GeV.

In PBH sterile neutrinogenesis, a sparse population of extremely high energy $\nu_s$s are produced.
We are interested in scenario where this population stays intact, and does not significantly turn into active neutrinos. Thus we consider the loss of these energetic $\nu_s$ into active neutrinos via their flavor oscillations using the same Boltzmann equation. 
We can utilize the same Boltzmann equation of Eq.~\eqref{eq:Boltzmann}, valid for temperatures significantly below 100 GeV, because as shown below the maximum of the $\nu_s$ loss rate occurs at these temperatures and we neglect inelastic scattering in the more general Boltzmann equation (see e.g. Ref.~\cite{Abazajian:2001nj}). This approximation simplifies the computation of the sterile neutrino density evolution, and typically introduces tolerably small errors~\cite{Dolgov:2000ew}. In the following we are interested in the disappearance of high energy sterile neutrinos and we show the regime in which the total interaction rate $\Gamma/H<1$ is small. Scattering, whether it entails a sterile-active transition and/or significant momentum transfer, also results in losing a high energy sterile neutrino. Therefore, for the purposes of computing the parameter space of no significant interactions and no thermalization, we can safely neglect the momentum exchange in the full Boltzmann equation.

Active neutrinos produced in the PBH evaporation rapidly thermalize at a much lower $T_{\rm RH}$ or $T_{\rm evap}$ temperature, which we require to be less than $0.1 T_{\rm PBH}$ for consistency.
The  energy of the thermalized $\nu_a$s (initially of the order of $T_{\rm evap}$ or $T_{\rm RH}$)  is much lower than the energy of the $\nu_s$  population (initially close to $T_{\rm PBH}$) so their abundance at high energy is exponentially suppressed. We set $f_{\nu_a}=0$ for high momentum interactions in Eq.~\eqref{eq:Boltzmann}.
Neglecting inelastic scattering and changing variables from $t$ to $T$, the $f_{\nu_s}$ term in Eq.~\eqref{eq:Boltzmann}, leads to
\begin{equation}
\label{eq:dfsdt}
    T\frac{df_{\nu_s}}{dT} \simeq \frac{1}{4H}\sin^2(2\theta_m)d_a \epsilon G_F^2 T^5 f_{\nu_s} = \frac{\Gamma_{\rm tot}}{H} f_{\rm \nu_s}~,
\end{equation}
which shows that  the high energy $\nu_s$ population distribution decreases with decreasing temperature at the total interaction rate $\Gamma_{\rm tot}$. Here $G_F \simeq 1.166 \times 10^{-5}$~GeV$^{-2}$ is the Fermi constant, $d_a$ is a flavor-dependent parameter close to 1
 ($d_a = 1.27$ if the corresponding lepton species is in the plasma and $d_a = 0.92$ otherwise), $f_{\nu,s}$ is the $\nu_s$ momentum distribution function, and $\sin^2 2\theta_m$ is the active-sterile mixing in  medium which has density (due to particle-antiparticle asymmetries) and thermal contributions from the plasma. 
As mentioned earlier, we do not assume presence of a significant lepton asymmetry, thus the contribution of asymmetries is negligible. Then, the in medium mixing in  Eqs.~\eqref{eq:Boltzmann} and~\eqref{eq:dfsdt} is  
\begin{equation}
\label{eq:mattermix}
    \sin^2(2\theta_m) = \frac{\sin^2(2\theta)}{\sin^2(2\theta) +
    \left[\cos(2\theta)-\dfrac{2\epsilon T}{m_s^2} V_T\right]^2}~.
\end{equation}
Below the electroweak scale, the thermal potential takes the form
\begin{equation}
    V_T \simeq -1 \times 10^{-8} \textrm{ GeV} \epsilon \left(\frac{T}{\textrm{GeV}}\right)^{5}~,
\end{equation}
if the charged lepton species corresponding to the active neutrino that mixes with the $\nu_s$ is populated. The high energy $\nu_s$ population will oscillate into active neutrinos if the rate $\Gamma_{\rm tot}$ is faster than the Hubble rate $H$.

We take the thermalization condition to be $\Gamma_{\rm tot}/H > 1$ at its maximum which corresponds to $|T df_{\nu_s}/dT|>1$ in Eq.~\eqref{eq:dfsdt}. At low temperatures, $|2\epsilon T V_T/m_s^2|\ll 1$ and $\Gamma_{\rm tot}/H \propto T^3$. When $|2\epsilon T V_T/m_s^2| > 1$, it dominates the denominator of Eq.~\eqref{eq:mattermix} so that $\Gamma_{\rm tot}/H \propto T^{-9}$. Above the electroweak scale, both the numerator of the interaction rate and thermal potential is proportional to $T$~\cite{Kuznetsov:2011rk,Kuznetsov:2013tta}, resulting in $\Gamma_{\rm tot}/H \propto T^{-5}$. Therefore there is only one maximum of $\Gamma_{\rm tot}/H$. Taking the derivative $d(\Gamma_{\rm tot}/H)/dT=0$, this maximum occurs when $|2\epsilon T V_T/m_s^2| \simeq 1/3$. Using the average value in Eq.~\eqref{eq:average-epsilon} of $\epsilon$ for $\nu_s$s at evaporation,
$\langle \epsilon\rangle = 6.3~(T_{\rm PBH}/T_{\rm evap})$, redshifted to the temperature at which the maximum of $\Gamma/H$, we find that this temperature is
\begin{equation}
    T_{\rm max,s} \simeq 810\textrm{ keV} \left(\frac{m_s}{\textrm{keV}}\right)^{1/3}\left(\frac{10^8{\rm g}}{M_{\rm PBH}}\right)^{1/6}\left(\frac{10.75}{g_*(T_{\rm max,s})}\right)^{1/12}~.
\end{equation}
This temperature is much lower than the $T_{\rm max}$ at which the DW production rate is maximum, given in 
Eq.~\eqref{eq:tmax}, due to the high momentum of the $\nu_s$s. We checked that the characteristic center of mass energy at the maximum of $\Gamma_{\rm tot}/H$,
\begin{equation}
\label{eq:ecmmax}
    E_{\rm CM} \simeq 6 \sqrt{\frac{T_{\rm PBH}}{T_{\rm RH}}}  T_{\rm max,s} \simeq 1.5 ~\textrm{GeV} \left(\frac{m_s}{\textrm{keV}}\right)^{1/3} \left(\frac{M_{\rm BH}}{10^8 g}\right)^{1/12} \lesssim 10^2~ {\rm GeV}~,
\end{equation}
is below the electroweak scale. Thus the propagator of gauge bosons exchanged in the $s-$channel is dominated by the boson mass.  The same is true for the  $t-$channel  exchange of weak gauge bosons, since  $\sqrt{t} \leq \sqrt{s} =E_{\rm CM}$ and 
therefore using the sub-electroweak form of the interaction rate and of the thermal potential is justified.
The thermalization condition $\Gamma_{\rm tot}/H > 1$ implies 
\begin{equation}
\label{eq:thermalization2}
    \sin^2 2\theta \geq 
    5 \times10^{-6} \left(\frac{\textrm{keV}}{m_s}\right) \left(\frac{g_*(T_{\rm max,s})}{10.75}\right)^{1/2}~,
\end{equation}
which is the same thermalization condition as Eq.~\eqref{eq:thermalization} for the DW mechanism. We see that both the thermalization of $\nu_s$s at low momenta and the scattering away of high energy $\nu_s$s occur for 
the same mixing angles even though they may happen at different temperatures. 
The reason is that  this condition does not depend on $\epsilon$ (since $\Gamma_{\rm tot}/H \propto \epsilon T_{\rm max}^3$ close to the maximum, and $T_{\rm max} \propto 1/\epsilon^{1/3}$).  Therefore, no additional condition on the mixing needs to be imposed by this consideration.

The conditions in the previous  section, Sec.~\ref{ssec:subdominant-DW}, for PBH sterile neutrinogenesis to  dominate over DW $\nu_s$ production, given in Eqs.~\eqref{eq:sinRstd}, ensure that the mixings we consider are below the thermalization range in Eq.~\eqref{eq:thermalization2} for all values of $T_{\rm evap}$ or $T_{\rm RH}$  that are $ < 5 \times 10^{15}~{\rm GeV}/ f_{\rm evap}$, even in the PD case, for which $f_{\rm evap}=1$,  when $T_{\rm RH} > T_{\rm max}$ (see Figs.~\ref{fig:Raddom} and~\ref{fig:PBHdom}). In the PD case when $T_{\rm RH} < T_{\rm max}$ (LRT)  the limit in Eq.~\eqref{eq:sinMLRT} implies $T_{\rm RH} > 0.2 (m_s/{\rm keV})^{3/8}$ (which e.g. for $T_{\rm RH}=5$~MeV is fulfilled for $m_s < 5$~MeV) and we do not have a region of interest outside this range (see Fig.~\ref{fig:PBHdom}).

\section{Evaporation in the radiation-dominated epoch (RD case)}
\label{sec:rdom}
 
\begin{figure}
    \centering
\includegraphics[width=0.7\textwidth]{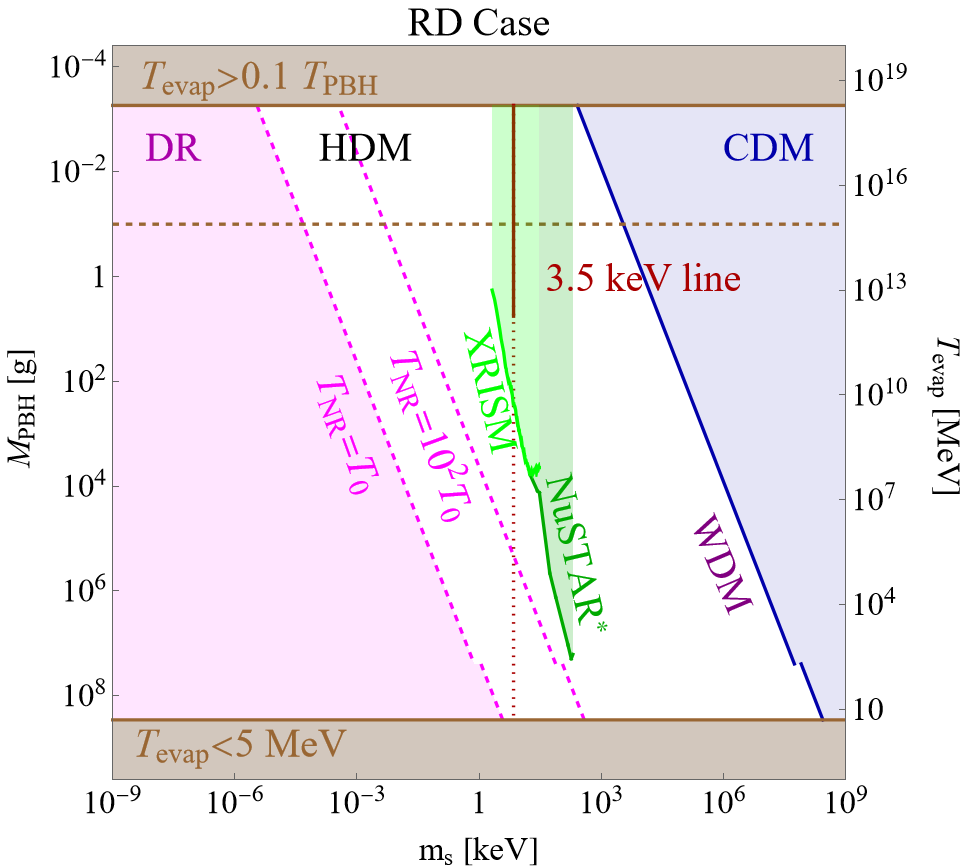}
    \caption{
    Regions and bounds  in the $M_{\rm PBH}$ (left axis) or  $T_{\rm evap}$ (right axis) versus mass $m_s$  for $\nu_s$s produced by  PBH  evaporation  in the radiation dominated epoch (the RD case). The upper and lower brown bands are excluded by the requirements on $T_{\rm evap}$ stated in text of Section~\ref{sec:evaporation}. Dotted magenta lines indicate where $\nu_s$s become non-relativistic at present, $T_{\rm NR}=T_0$, and at redshift 100, $T_{\rm NR}= 100~ T_0$. Only $\nu_s$s in the blue region can account for all of the DM (i.e. $f_s=1$) and be either WDM along the blue solid boundary line or CDM above it.  Below the blue line $\nu_s$s can be either HDM, in the white region, and thus they must have $f_s <0.1$, or DR (relativistic at present) in the magenta region.  Also indicated are regions in which $\nu_s$s from PBH evaporation can account for the putative $3.5$ keV line (brown vertical line), a potential future  X-ray signal in XRISM (lighter green region), or a possible improved NuSTAR-like search we call   NuSTAR$^*$ (darker green region). The region above the horizontal dashed brown line is restricted by the upper bound on the energy scale of inflation $\rho_{\rm inf} < (1.6 \times 10^{16} \textrm{ GeV})^4$~\cite{Planck:2018jri} (see text).
    }
    \label{fig:Raddom}
\end{figure}

In Fig.~\ref{fig:Raddom} we display all the regions of interest and limits in the $M_{\rm PBH}$ (left axis) or  $T_{\rm evap}$ (right axis) versus mass $m_s$ space for $\nu_s$s produced by PBH sterile neutrinogenesis in the RD scenario. As mentioned in Sec.~\ref{sec:PBH}, we exclude  evaporation temperatures $T_{\rm evap} > 0.1~T_{\rm PBH}$ and $T_{\rm evap}< 5~{\rm MeV}$ due to consistency considerations and limits imposed by BBN respectively, as shown in the upper and lower brown bands of Fig.~\ref{fig:Raddom}.

The $\nu_s$ abundance
in the RD case is suppressed by the PBH fraction of the DM at evaporation, $f_{\rm evap} <1$ in
Eq.~\eqref{eq:fevaprad},
which makes possible to have $\nu_s$s comprise all of DM and be either WDM or CDM.  As we explain below, the Lyman-$\alpha$ limit allows $f_s=1$ in the blue region of Fig.~\ref{fig:Raddom}, at and above the blue line for WDM and CDM, respectively.

Below the blue line, $m_s$ is below the Lyman-$\alpha$ limit and can thus only be either HDM (in the white region of Fig.~\ref{fig:Raddom}) or dark radiation (DR), i.e. relativistic at present, in the magenta region of the same figure. In the DR region sterile neutrinos make a contribution $\Delta N_{\rm eff} \simeq 6.8\times10^{-2} f_{\rm evap}$ (see e.g. Eq.~(22) of~\cite{Hooper:2019gtx}) to the effective number of neutrino species. This contribution is much smaller than current upper limits but could be probed by CMB-S4, whose expected sensitivity is $\Delta N_{\rm eff} \leq 0.06$~\cite{Abazajian:2019eic}. 

Because the $\nu_s$s produced by PBH sterile neutrinogenesis are highly energetic they can comprise DR even in the keV-mass range. 
For the same reason they become non-relativistic at a temperature $T_{\rm NR}$ if their mass is  much larger than for other production mechanisms,
\begin{equation}
\label{eq:NR}
  m_{s} >\langle p\left( T_{\rm NR}\right)  \rangle=\langle\epsilon\rangle T_{\rm NR}\left[\dfrac{g_{*}\left( T_{\rm NR} \right) }{g_*\left( T_{\rm evap}\right) }\right] ^{{1}/{3}} \simeq 2 \times 10^{7} T_{\rm NR} 
  \left[\dfrac{5~{\rm MeV}}{T_{\rm evap}}\right]^{1/3}
  \left[\dfrac{10.75}{g_*(T_{\rm evap})}\right]^{1/6},
\end{equation}
where $\langle\epsilon\rangle $ is the average temperature-scaled $\nu_s$ momentum at evaporation given in Eq.~\eqref{eq:average-epsilon}. Thus, the temperature at which a $\nu_s$ with mass $m_s$ becomes non-relativistic is
\begin{equation}
\label{eq:TNR-RD}
  T_{\rm NR} \simeq 10^{-4}\textrm{eV} 
  \left(\frac{m_s}{\rm keV}\right)\left(\frac{10^8 {\rm g}}{M_{\rm PBH}}\right)^{1/2}~.
\end{equation}
In Fig.~\eqref{fig:Raddom}, the DR magenta  region corresponds to $T_{\rm NR}<T_0$. The figure also shows  the line of  $T_{\rm NR}< 100~ T_0$ at which $\nu_s$s become non-relativistic at redshift $z=100$.  Below this line, $\nu_s$s are non-relativistic early enough as to be part of the DM in  different structures in the Universe, such as galaxy clusters, and thus decay and produce  X-ray or $\gamma$-ray lines.
The approximate lower limit on $T_{\rm NR}$ for $\nu_s$s to be WDM (or CDM) is $T_{\rm NR}> {\rm keV}$ and approximately coincide with the Lyman-$\alpha$ we limit present next.  

\subsection{Lyman-$\alpha$ forest limit in the RD case}

From Fig.~6 of Ref.~\cite{Baur:2017stq} Lyman-$\alpha$ (Ly-$\alpha$) forest limits imply that a DM fermion with a thermal spectrum (i.e. a ``thermal fermion")  is disfavored as the sole DM constituent at the $2-\sigma$ level if its mass is $m_{\rm therm} \leq 3$ keV (and if its mass is $<1$ keV, it can only account for  fraction
$\lesssim 0.1$ of the DM). A thermal fermion of mass close to 3 keV would constitute WDM and heavier ones would be CDM.

This bound can be extended to $\nu_s$s using the following
relation initially given in Ref.~\cite{Baur:2017stq} and later modified in Ref.~\cite{Gelmini:2019wfp} to include the average scaled momentum $\langle \epsilon\rangle$ dependence 
\begin{align}
\label{eq:massrelation}
    m_s \simeq &~ 
    4 \textrm{ keV} \left(\frac{\langle \epsilon\rangle}{3.15}\right) \left(\frac{m_{\rm therm}}{\textrm{ keV}}\right)^{4/3} \left(\frac{10.75}{g_*(T_{\rm evap})}\right)^{1/3} ~.
\end{align}
Due to the high  $\langle \epsilon\rangle$ value at evaporation given in Eq.~\eqref{eq:average-epsilon}, the Lyman-$\alpha$ bound applies to heavier particles,
\begin{equation}      
\label{eq:lymanR}
    m_s \geq 3~{\rm keV} \times 10^8\left( \frac{5 \,\rm MeV}{T_{\rm evap}} \right)^{1/3}\left( \frac{10.75}{g_\ast(T_{\rm evap})}\right)^{1/6}~. 
\end{equation}
Using this lower limit on the present DM fraction in $\nu_s$s, $f_s$ that is given in Eq.~\eqref{eq:fnr-RD}, we find that one can always have $f_s=1$ above the Lyman-$\alpha$ limit, by considering the appropriate fraction of the total energy density associated with PBHs at evaporation,  $f_{\rm evap}<1$. 
The Lyman-$\alpha$ limit in  Eq.~\eqref{eq:lymanR} is shown as the solid blue line in Fig.~\ref{fig:Raddom}.
As we mentioned, close to this line, $\nu_s$ are WDM and in the blue region above it they are CDM.  
For example, along the boundary line where $\nu_s$s are WDM, in order to have $f_s=1$ (i.e. having all of the DM)  the required PBH density fraction at evaporation is
\begin{equation}
    \label{eq:fevapWDM}
    f_{\rm evap}\simeq 
    4 \times 10^{-3}\left(  \frac{g_*(T_{\rm evap})}{10.75}\right)^{1/3}~.
\end{equation}
We highlight that the WDM line extends from  3 MeV to 0.3 TeV in $\nu_s$ mass range, very distinct from the typical keV mass range expected for $\nu_s$ WDM in other production scenarios.

Let us clarify that the Lyman-$\alpha$ constraint  given in Eq.~\eqref{eq:lymanR} in terms of the minimum thermal mass $m_{\rm therm}$ coincides with the often used WDM limit given in terms of 
the present-day characteristic DM speed $v_0$. For a thermal WDM particle with speed distribution $f_{\rm WDM}(v) = (e^{v/v_0}+1)^{-1}$, the present  characteristic speed is~\cite{Bode:2000gq}
\begin{equation}
\label{eq:v0}
    v_0 \simeq 4\times 10^{-8}\left[ \left( \frac{ \Omega_{\rm therm} h^2}{0.12}\right) \left( \frac{\rm keV}{m_{\rm therm}}\right) ^4\right]^{1/3}.
\end{equation} 
For $\nu_s$ produced through PBH sterile neutrinogenesis, the characteristic speed at present is
\begin{equation}
\label{eq:vs0}
    v_{s,0}= \frac{\langle p(T_0)\rangle}{m_s} \simeq 1.64\times10^{-7} \left(\frac{\langle \epsilon \rangle}{3.15}\right)\left(\frac{\rm keV}{m_s}\right) \left(\frac{10.75}{g_s(T_{\rm evap})}\right)^{1/3}~.
\end{equation}
 Equating Eqs.~\eqref{eq:v0} and~\eqref{eq:vs0}, we get the relation in Eq.~\eqref{eq:massrelation} between $m_s$ and the thermal fermion mass. 
 We believe that a factor of $3$ discrepancy found in  Ref.~\cite{Baldes:2020nuv} between the limit using $v_0$ and a more elaborate method of obtaining the WDM limit, stems from taking $v_0$ to be the average speed instead of the characteristic speed for a thermal fermion, which  differ by a factor of $3.15$ for a thermal Fermi-Dirac distribution. 

\subsection{Possible X-ray line signals in the RD case}
\label{ssec:radxray}

The presence of $\nu_s$s as constituents of DM in galaxies and galaxy clusters can be probed through their decays, especially into photons, and also through their impact on structure formation. In the keV-mass range, $\nu_s$s generate X-ray lines through $\nu_s\xrightarrow{} \nu_a \gamma$, which could have already been detected if the putative 3.5 keV line signal is confirmed, or could be detected in the future by  X-ray experiments such as a possible future  NuSTAR-like experiment we call ``NuSTAR$^*$", with twice the sensitivity of the Nuclear Spectroscopic Telescope Array (NuSTAR)~\cite{Ng:2019gch} in the $\sim 10-200$~keV band, 
or the recently launched  
X-Ray Imaging and Spectroscopy Mission
(XRISM) satellite~\cite{XRISMScienceTeam:2020rvx} (sensitive to X-ray energy band of $\sim 0.4-15$~keV). 

The non-observation of such X-ray or $\gamma$-ray lines sets limits on the mixing angle $\sin^2 2\theta$ for each $m_s$ value (e.g. they impose that $\nu_s$ decays into $e^+ e^-$ pairs make a very small contribution to the persistent 511~keV signal observed from the Galactic central region, as we comment in App.~\ref{appsec:511kev}). However, since PBH evaporation produces $\nu_s$s independently of their mixing angle, such limits do not independently reject any region of the $m_s- M_{\rm PBH}$ parameter space.  

Instead, the observation of an X-ray signal may be due to $\nu_s$s with active-sterile mixing below the limit in Eq.~\eqref{eq:sinRstd} for which $\nu_s$s are dominantly produced by PBH sterile neutrinogenesis, i.e. $f_s>f_{s,osc}$ (see Sec.~\ref{ssec:abundance}). In this case, we determine the region in the $m_s-M_{\rm PBH}$  parameter space from which such a signal can stem.    
In Fig.~\ref{fig:Raddom}, we show the regions (green) corresponding to a possible observation of an X-ray line in the aforementioned experiments and a vertical brown line for the putative 3.5 keV signal. Each possible observation of an X-ray emission line at an energy $E_\gamma$ in these experiments can stem from a vertical line within that green region at $m_s = 2 E_\gamma$. 

The 3.5 keV line requires $m_s=7.1~{\rm keV}$, and for $f_s=1$ a mixing $ \sin^2(2\theta) = 5\times 10^{-11} $~\cite{Bulbul:2014sua,Boyarsky:2014jta}. However, $\nu_s$s with this mass produced in the RD case  are HDM and thus have $f_s<0.1$.  Since the observed intensity of the line  depends on the product of the decay rate $\propto  \sin^2(\theta)$ times the density $\propto f_s$, $\nu_s$s produced in the RD case  can explain the 3.5 keV line signal if $f_s \sin^2(2\theta) = 5\times 10^{-11}$. Hence, the required mixing  increases inversely with density as $f_s^{-1}$. 
Using this required mixing and $m_s=7.1~{\rm keV}$ in Eq.~\eqref{eq:fsdmosc} for $f_{s,osc}$, the condition $f_s>f_{s,osc}$ for $\nu_s$s to be dominantly produced by PBH sterile neutrinogenesis, becomes
\begin{equation}
    f_s>1.3 \times 10^{-2}\left( \frac{5\times 10^{-11}}{10^{-10}f_s}\right)\left( \frac{30}{g_*(T_{\rm max})} \right)^{3/2}, 
\end{equation}
thus
\begin{equation}
     f_s^2>6.5\times 10^{-3}\left( \frac{30}{g_*(T_{\rm max})} \right)^{3/2},
\end{equation}
which gives a lower bound on the density fraction, $f_s>0.032$, when $g_*(T_{\rm max})=106.75$. 
Since $f_s$, as shown in Eq.~\eqref{eq:fnr-RD}, is proportional to the product $f_{\rm evap} (T_{\rm evap}/{\rm 5~MeV})^{1/3}$,  this lower bound on $f_s$ translates into a lower bound on $T_{\rm evap}$,
\begin{equation}
    T_{\rm evap}>
    1.4 \times 10^{12}~{\rm MeV}\left( \frac{1}{f_{\rm evap}^3} \right)\left( \frac{g_*(T_{\rm evap})}{106.75}\right)^{1/2}.
\end{equation}
Therefore, the condition of the RD case $f_{\rm evap} < 1$ results in the lower limit
$T_{\rm evap}>1.4\times 10^{12}$~MeV, which indicates the lowest point of the  vertical brown line in Fig.~\ref{fig:Raddom}.
The upper limit of this line is given by the consistency condition $T_{\rm evap} <0.1~ T_{\rm PBH} \simeq 1.9~\times10^{18}~{\rm MeV}$ discussed in Sec.~\ref{sec:evaporation}, corresponding to the lower limit of the upper excluded brown band region in Fig.~\ref{fig:Raddom}.

In the same manner we determine the range of parameters in which a possible future signal in XRISM or an improved NuSTAR-like experiment, NuSTAR$^*$, could be due to $\nu_s$s produced through PBH sterile neutrinogenesis. For each $m_s$ and assumed mixing we find a vertical line, similar to the brown line shown for the 3.5 keV signal, and the union of all these lines constitutes the green region. 
For the assumed mixing value of the signal, we take the estimated reach of XRISM given in Fig.~S8 of Ref.~\cite{Dessert:2023vyl}  and for
NuSTAR$^*$, we use the limits in Fig.~6 of Ref.~\cite{Roach:2022lgo}, assumed to be improved by a factor of two. The upper limit on $T_{\rm evap}$
stays unchanged, and  the lower limit stems from the $f_s>f_{s,osc}$ condition. 
The parameter region these two experiments can probe is shown in Fig.~\ref{fig:Raddom} in light green for XRISM and dark green for NuSTAR$^*$, respectively.  Since the sterile neutrino density $f_s> 10^{-4}$ is moderately large in these regions, the mixing angle required to produce a detectable X-ray signal (a product of the density and mixing angle) results in a $\nu_s$ lifetime always larger than the age of the Universe,  $\tau \gg t_U$. Thus, the decays in this regime do not significantly reduce the sterile neutrino abundance.

Considering that the possible detected X-ray signal comes from DM in galaxy clusters, $\nu_s$s need to non-relativistic early enough to ensure that they are bound into structures in the Universe. Notice that the (green) regions of interest are
above $T_{\rm NR}= 100~ T_0=2.3\times 10^{-2}~{\rm eV}$ (dashed magenta line in Fig.~\ref{fig:Raddom}), corresponding to $\nu_s$s that became non-relativistic at
redshifts larger than $z=100$. 

\section{Evaporation in the PBH matter-dominated epoch (PD case)} 
\label{sec:mdom}

\begin{figure}
    \centering
    \includegraphics[width=0.7\textwidth]{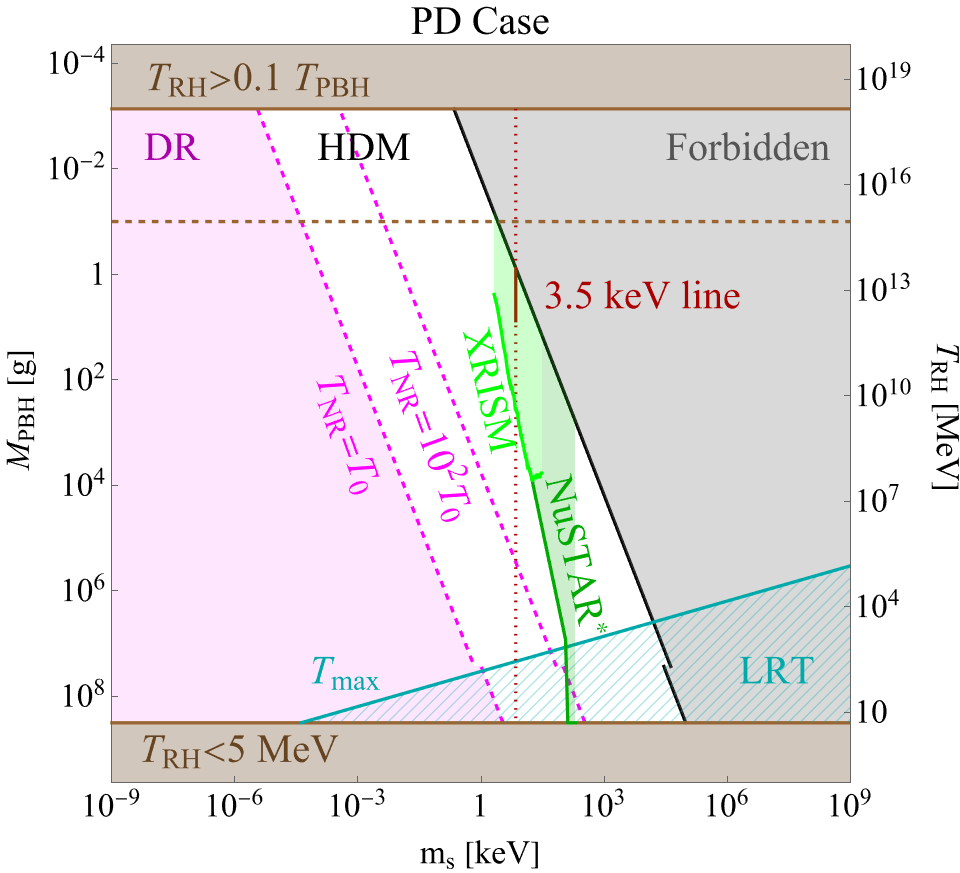}
    \caption{
    Regions of interest and bounds in the $M_{\rm PBH}$ (left axis) or $T_{\rm RH}$ (right axis) versus mass $m_s$ for $\nu_s$s produced in the evaporation of PBHs that matter-dominate the Universe (PD case). The upper and lower brown bands are excluded by the requirements on  $T_{\rm RH}$. Dotted magenta lines indicate where $\nu_s$s become non-relativistic at the indicated temperature. In the gray region, $\nu_s$ above the Ly-$\alpha$ limit would be overabundant and are thus forbidden. Below the black line,  $\nu_s$ below the Ly-$\alpha$ limit have $f_s<0.1$, thus they are either HDM (white region) or DR relativistic at present (magenta region). Regions in which the putative 3.5 keV signal (vertical brown line) or a potential future X-ray signal in XRISM (lighter green) or an improved NuSTAR-like search  we call NuSTAR$^*$ could be due to $\nu_s$s from PBH neutrinogenesis are indicated. In the LRT (hatched in cyan), the production via oscillations is suppressed because $T_{\rm RH}$ is below $T_{\rm max}$ (cyan line). The region above the horizontal dashed brown line is restricted by the upper bound on the energy scale of inflation $\rho_{\rm inf} < (1.6 \times 10^{16} \textrm{ GeV})^4$~\cite{Planck:2018jri} (see text).}
    
    \label{fig:PBHdom}
\end{figure}

In Fig.~\ref{fig:PBHdom} we show the limits and regions of interest in the $M_{\rm PBH}$ (left axis) or  $T_{\rm RH}$ (right axis) versus mass $m_s$ space for $\nu_s$s produced by PBH sterile neutrinogenesis in the PD scenario. As mentioned in Sec.~\ref{sec:PBH}, we exclude regions of reheating temperatures $T_{\rm RH} > 0.1~T_{\rm PBH}$ and $T_{\rm RH}<  5$~ MeV due to consistency considerations and limits imposed by BBN, respectively. This is shown in the upper and lower brown bands of Fig.~\ref{fig:PBHdom}. 

We present for the PD case similar limits and regions of interest as in the previous Sec.~\ref{sec:rdom} for the RD case. The main difference  is that in the PD case the PBH fraction of the DM at evaporation is $f_{\rm evap}=1$ and thus the present DM fraction in $\nu_s$, $f_s$ in Eq.~\eqref{eq:fnr-RD}, is a fixed function of the $M_{\rm PBH}$ and $m_s$. 

The upper limits on the active-sterile mixing to guarantee that the DW produced population of $\nu_s$s is subdominant, as explained in Sec.~\ref{ssec:subdominant-DW}, depend of having  $T_{\rm RH} > T_{\rm max}$  or $T_{\rm RH} < T_{\rm max}$ (which is an LRT scenario, corresponding to the region hatched in cyan in Fig.~\ref{fig:PBHdom}). The limits are given in Eq.~\eqref{eq:sinRstd} and in Eq.~\eqref{eq:sinMLRT}, respectively.
The lower limit on $m_s$ for $\nu_s$s to be non-relativistic at the temperature $T_{\rm NR}$ is the same as in Eq.~\eqref{eq:NR} with $T_{\rm evap}$ replaced by $T_{\rm RH}$. The temperature $T_{\rm NR}$ at which a $\nu_s$ with given mass $m_s$ becomes non-relativistic is the same as  Eq.~\eqref{eq:TNR-RD}. Similarly to the RD case, for the PD case the DR region (magenta) for which $\nu_s$s are still relativistic at present, i.e. $T_{\rm NR} <T_0$,  and  the $T_{\rm NR} = 100 T_0$ line
(dashed magenta), are shown in Fig.~\ref{fig:PBHdom}.

\subsection{Lyman-$\alpha$ forest  limit in the PD case}
\label{ssec:mlyman}

To obtain limit from Lyman-$\alpha$ forest for the PD case,  $T_{\rm evap}$ needs to be replaced by $T_{\rm RH}$ in the relation between $m_s$ and $m_{\rm therm}$ in Eq.~\eqref{eq:massrelation} as well as in the lower limit imposed on $m_s$ in  Eq.~\eqref{eq:lymanR}. Then, we use this limit in $f_s$ as in the RD case. 
The main difference from the RD case is that now $f_s$
is a fixed function of $M_{\rm PBH}$ and $m_s$, because $f_{\rm evap}=$ 1. As we will explain, this implies
that $\nu_s$s cannot constitute all of DM, i.e  $f_s$ cannot be unity in any region of the whole parameter space.

For masses at and above the Lyman-$\alpha$ limit  $\nu_s$s would be overabundant (with $f_s\gtrsim 95$) and thus they are forbidden (they are within the gray region in Fig.~\ref{fig:PBHdom}).
For $m_s$ values below the Lyman-$\alpha$ limit,  $\nu_s$s are either HDM and thus must have $f_s<0.1$ (in the white region) or DR (magenta region).  The condition $f_s<0.1$ using Eq.~\eqref{eq:fnr-RD} translates into an upper limit on $m_s$,
\begin{equation}
\label{eq:PD-HDM-allowed}
    m_s \leq 1 \times 10^{5} ~{\rm keV} \left( \frac{5~ {\rm MeV}}{T_{\rm RH}}\right) ^{1/3} \left( \frac{g_*( T_{\rm RH}) }{10.75}\right)^{{1}/{6}}. 
\end{equation}
This upper limit $f_s = 0.1$ is indicated with a  black line in Fig.~\ref{fig:PBHdom}. The gray region above it is forbidden either because $m_s$ is below the Ly-$\alpha$ limit in~Eq.~\eqref{eq:lymanR} (thus $\nu_s$s are HDM) and $f_s>0.1$, or because it is above it but $f_s$ would be necessarily much larger than 1.

\subsection{Possible X-ray line signals in the PD case}

For the putative 3.5 keV line in the PD case, we impose the same two conditions given for the RD case in the preceding section.
The  condition $f_s>f_{s,osc}$ gives the same lower bound $f_s>0.032$, or equivalently $T_{\rm RH}>1.4\times 10^{12}~{\rm MeV}$.  Note that this lower bound is much larger than $T_{\rm max}$ (given in Eq.~\eqref{eq:tmax}) so the brown vertical line is above the LRT region. 
While the lower limit of the brown line is the same as for the RD case, the upper limit is not. As discussed above, $\nu_s$s in the PD case are HDM, thus must have $f_s<0.1$, i.e.it extends to  the black line in Fig.~\ref{fig:PBHdom}, which for $m_s=7.1$~keV translates into the upper bound $T_{\rm RH}<4.16\times 10^{13}~{\rm MeV}$. 
We again show the range satisfying the two conditions in Fig.~\ref{fig:PBHdom} by the brown vertical line.

Similarly to the RD case, we also show in Fig.~\ref{fig:PBHdom} the regions in which mixing angles potentially probed by the X-ray experiments XRISM (lighter green) and  a possible NuSTAR$^*$ (darker green) could be due to PBH sterile neutrinogenesis. Lower projected sensitivity bounds for XRISM  and NuSTAR$^*$ regions are obtained in the same manner as in Sec.~\ref{ssec:radxray}, but the upper sensitivity bound of the green regions is given by $f_s<0.1$ (black line in Fig.~\ref{fig:PBHdom}). 
The possible NuSTAR$^*$ could probe the LRT region (Eq.~\eqref{eq:fsLRTosc}), where DW production is suppressed and we require $f_s>f_{s,osc}^{\rm LRT}$. This region  ends at the boundary of the lower brown band, due to the limit  $T_{\rm RH}>5~{\rm MeV}$ imposed by BBN.

Notice that the  $\nu_s$s in the green regions became non-relativistic close to $z=100$ ($T_{\rm NR}>100~T_0=2.3\times 10^{-2}~{\rm eV}$, corresponds to $T_{\rm RH}>4.8\times10^{5}~{\rm MeV}$). 

\subsection{Gravitational wave signals in the PD case}

The production of GWs associated with formation of PBHs 
has been studied in many contexts.
As scalar and tensor perturbations are coupled at non-linear level, induced GWs can be generated at second order in cosmological perturbation theory due to curvature fluctuations~\cite{Tomita:1967,Ananda:2006af,Baumann:2007zm,Saito:2010}.
Induced GWs have been associated with large primordial curvature perturbations preceding and giving rise to PBHs~(e.g.~\cite{Cai:2018dig}).

Subsequently to their production, while the PBHs are a subdominant population, fluctuations in their number density constitute isocurvature perturbations. 
Here, we consider GWs induced at second-order that are produced when the PBHs matter-dominate, converting their isocurvature perturbations to sizable curvature
perturbations. Then, density perturbations in matter are rapidly converted to radiation during near-instantaneous evaporation, inducing GWs~\cite{Inomata:2019zqy,Inomata:2020lmk,Papanikolaou:2020qtd,Domenech:2020ssp,Domenech:2021wkk}. These perturbations and the GW signal would be suppressed in the RD case by the small contribution of PBH to the energy density at evaporation, $f_{\rm evap} <1$, hence we consider them only in the context of PD. Here we make the simplifying assumption that the PBH mass spectrum is monochromatic, although in principle the GW spectrum from evaporating PBHs could be calculated even for a continuous distribution.

\begin{figure}
    \centering
\includegraphics[width=0.7\textwidth]{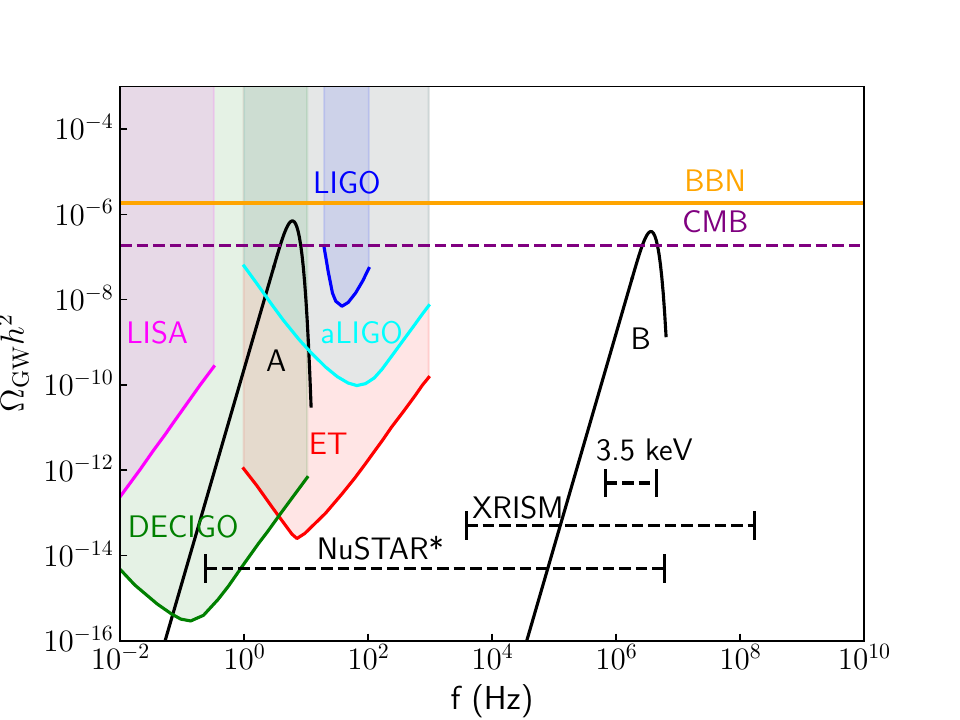}
    \caption{
    GW density induced in the PBH reheating scenario associated with sterile neutrinogenesis in terms of GW frequency. Target parameter space where coincidence signatures from decaying neutrinos from PBH sterile neutrinogenesis can result in X-ray signatures is also displayed. 
    The solid lines correspond to models A ($M_{\rm PBH}=2\times10^7$~g, $\beta=6\times10^{-9}$), and B ($M_{\rm PBH} = 1$~g, $\beta =8\times10^{-4}$). Also shown are the sensitivity curves for LISA, DECIGO, ET and Advanced LIGO GW experiments~\cite{Thrane:2013oya}, the 95\% C.L. limits on $\Delta N_{\rm eff}$ from BBN~\cite{Arbey:2021ysg}, and projections from $\Delta N_{\rm eff}$ of CMB-S4~\cite{Abazajian:2019eic}. 
    }
    \label{fig:EvaporatingGW}
\end{figure}

The present day energy density of the resulting GW background  follows a steep power law~\cite{Domenech:2021wkk}
\begin{equation}
\label{eq:gwreh}
    \Omega_{\rm GW, 0} = \Omega_{\rm GW, 0}^{\rm peak} \left(\frac{k}{k_{\rm UV}}\right)^{5} \Theta(k_{\rm UV}-k)~.
\end{equation}
This GW spectrum has a sharp peak at  the frequency $f_{\rm UV}$ corresponding to wavenumber $k_{\rm UV}$, the ultraviolet cut-off of the gravitational potential fluctuations power spectrum, 
\begin{equation}
\label{eq:fpeak}
    f_{\rm UV} = 1.7\times10^{3}\textrm{ Hz} \left(\frac{M_{\rm PBH}}{10^4 \rm g}\right)^{-5/6}~,
\end{equation}
 with a rapid decline and cutoff around $k \sim 2 k_{\rm UV}$~\cite{Domenech:2020ssp,Inomata:2020lmk}. The present density of GWs at the peak in relation to the present 
 radiation density $\Omega_{r,0} h^2=8\times10^{-5}$
 is given by~\cite{Domenech:2020ssp,Domenech:2021wkk}
\begin{align}
\begin{split}
\label{eq:omegapeak}
    \Omega_{\rm GW,0}^{\rm peak} h^2 = &~ 0.39 \frac{1}{49152\pi \sqrt{3}}\left(\frac{3}{2}\right)^{2/3}\Omega_{r,0}h^2 \left(\frac{k_{\rm UV}}{k_{\rm RH}}\right)^{17/3}\left(\frac{k_{\rm eq}}{k_{\rm UV}}\right)^8\left(\frac{g_{*}(T_{\rm RH})}{106.75}\right)^{-1/3} \\ = &~ 1.64\times10^{-6} \left(\frac{\gamma}{0.2}\right)^{7/9}\left(\frac{\beta}{10^{-8}}\right)^{16/3}\left(\frac{M_{\rm PBH}}{10^7 \rm g}\right)^{34/9}~.
\end{split}
\end{align}
Here $k_{\rm RH}$ and $k_{\rm eq}$ corresponds to a frequency inverse of the horizon scale at reheating and at matter-radiation equality, respectively.
In the second line, PBH formation from collapse of large primordial perturbations
described in Sec.~\ref{PBH-formation} is assumed. We show the induced GWs generated by two PD models, denoted A ($M_{\rm PBH}=2\times10^7$~g, $\beta=6\times10^{-9}$), and B ($M_{\rm PBH} = 1$~g, $\beta =8\times10^{-4}$)
in Fig.~\ref{fig:EvaporatingGW}, plotted against the sensitivities of existing and future GW experiments. Using Eq.~\eqref{eq:fpeak} to relate the peak frequency to the PBH mass, we show the spans in frequency space in which the X-ray observations (see green regions in Fig.~\ref{fig:PBHdom}) could be correlated with a coincident GW background.  Note that even for high frequencies in which GW detection is difficult, a large GW background could contribute significantly to additional effective number of neutrino species $\Delta N_{\rm eff}$~\cite{Caprini:2018mtu} and could be observable in the future, such as by CMB-S4~\cite{Abazajian:2019eic}.

\section{Conclusions}
\label{sec:conclusions}

We studied $\nu_s$s that minimally couple to the SM only through a mixing with an active neutrino $\nu_a$. These are well motivated particles beyond SM that are actively searched for experimentally and are considered prime WDM candidates. Our work demonstrates that if they are produced through the evaporation of PBHs, their astrophysical and cosmological characteristics as DM constituents are remarkably distinct compared to $\nu_s$s produced through all other mechanisms studied so far. PBH sterile neutrinogenesis depends only on the $\nu_s$ mass and thus $\nu_s$s can constitute all or a significant fraction of the DM even if their mixing with active species is very small.

In the keV mass range, $\nu_s$s are usually considered good WDM candidates, but in our scenario they are HDM. They could, however be WDM in the much heavier 0.3 MeV to 0.3 TeV mass range instead, if PBHs evaporate in the radiation-dominated era in which they are formed (the RD case). In this mass range, $\nu_s$s could account for all of the DM, and as WDM would have a distinctively significantly hotter spectrum than in other scenarios. In the RD case, $\nu_s$s could also constitute CDM and account for all of the DM for masses between 1 MeV to 1 TeV.   
 
If PBHs matter-dominated the Universe before evaporating (the PD case), nearly instantaneous reheating at evaporation yields GWs with a characteristic frequency spectrum. This stochastic background could be detected in GW detectors or other cosmological probes, and be correlated with an X-ray line signal due to $\nu_s$ decays in galaxies or galaxy clusters, either in upcoming instruments or if the putative 3.5 keV possibly observed signal is confirmed. The detection of these coincident signals would constitute an identifying signature of our scenario.

Our study opens new directions to explore fundamental aspects of $\nu_s$s and connects them with different fields of research.

\section*{Acknowledgements}
\addcontentsline{toc}{section}{Acknowledgments}

G.G. and M.C. were supported in part by the U.S. Department of
Energy (DOE) Grant No. DE-SC0009937 and M.C. also by the UC Southern California Hub, with funding from the UC National Laboratories division of the University of California Office of the President. P.L. was supported by Grant Korea NRF2019R1C1C1010050.
V.T. acknowledges support by the World Premier International Research Center Initiative (WPI), MEXT, Japan and JSPS KAKENHI grant No. 23K13109.

\appendix

\section{Sterile neutrino production from active-sterile oscillations}
\label{app:osc}

Sterile neutrinos $\nu_s$s are produced in the early Universe through their mixing with the active neutrino species $\nu_a$ via active-sterile flavor oscillations and interactions that act as measurements. In this Dodelson-Widrow (DW)~\cite{Dodelson:1993je} mechanism, the production can be computed to a good approximation  using the Boltzmann equation~\cite{Kolb:1990vq,Abazajian:2001nj} 
\begin{equation}
\label{eq:Boltzmann}
    \frac{df_{\nu_s}}{dt}(p, t) \simeq \frac{1}{4}\sin^2(2\theta_m) G_F^2 \epsilon T^5 \left[f_{\nu_a} (p, t)(1-f_{\nu_s}(p, t)) - f_{\nu_s}(p, t)(1-f_{\nu_a}(p, t))\right]~,
\end{equation}
where $\sin^2 2\theta_m$ is the active-sterile mixing in matter (given in Eq.~\eqref{eq:mattermix}), $G_F$ is the Fermi constant, $\epsilon=p/T$ is the temperature scaled dimensionless momentum, and $f_{\nu_s}$, $f_{\nu_a}$ are the phase-space density distribution functions of the sterile and active neutrino populations, respectively. In this equation, momentum transfer is neglected, which results in small errors~\cite{Dolgov:2000ew}.

In contrast to PBH sterile neutrinogenesis, which produces $\nu_s$s irrespective of their couplings, the DW production from oscillations scales linearly with $\sin^2 2\theta$~\cite{Gelmini:2019wfp}
\begin{equation}
\label{eq:fsdmosc}
    f_{s,\textrm{osc}} =2.7\times10^{-4} \left(\frac{\sin^2 2\theta}{10^{-10}}\right)\left(\frac{m_s}{\textrm{keV}}\right)^2 \left(\frac{30}{g_*(T _{\max} )}\right)^{3/2}~.
\end{equation}
The DW production rate  has a sharp peak at the temperature~\cite{Dodelson:1993je}
\begin{equation}
\label{eq:tmax}
T _{\max}\simeq 133\textrm{ MeV}\left( \dfrac{m_s}{\rm keV}\right)^{1/3}.
\end{equation}
For heavy $\nu_s$ with masses $m_s \gtrsim 100$ MeV, 
the abundance of $\nu_s$ is increased by about one order of magnitude due to a resonance induced by a sign change in the thermal potential above the electroweak scale~\cite{Alonso-Alvarez:2022uxp}. Although this affects the precise conditions for PBH sterile neutrinogenesis domination, it does not significantly alter our results even in the affected region.

If the reheating temperature is low, $T_{\rm RH}<T_{\rm max}$, instead, the low reheating temperature (LRT) model should be used~\cite{Gelmini:2019wfp},
\begin{equation}
\label{eq:fsLRTosc}
f^{\rm LRT}_{osc} \simeq 1\times 10^{-7}\left( \dfrac{\sin ^{2}2\theta }{10^{-10}}\right) \left( \dfrac{m_s}{\rm keV}\right) \left( \dfrac{T_{\rm RH}}{5~{\rm MeV}}\right)^{3}~.
\end{equation}

\section{Sterile neutrino decays}

 The decay of $\nu_s$s due to their mixing with active neutrinos play an important role in their phenomenology, including their relic abundance, potential signals as well as constraints. Depending on the $\nu_s$ mass and mixing, different $\nu_s$ decay channels become relevant~\cite{Barger:1995ty,Atre:2009rg,Fuller:2011qy}.
The dominant decay channel for $m_s$ below the pion rest mass ($m_{\pi^0} = 135$~MeV, $m_{\pi^{\pm}} = 139.57$ MeV)  is into three active neutrinos with decay rate~\cite{Boehm:1987fc,Barger:1995ty}
\begin{equation}
\label{eq:decay}
    \Gamma_{3\nu} = \frac{G_F^2}{768\pi^3} m_s^5 \sin^2 2\theta \simeq 8.7 \times 10^{-31}~\textrm{s}~\Big(\dfrac{\sin^2 2\theta}{10^{-10}}\Big)\Big(\dfrac{m_s}{1~\textrm{keV}}\Big)^5~.
\end{equation}
We can compare the corresponding decay rate to the lifetime of the Universe $t_0=13.6$ Gyr
\begin{equation}
\label{eq:gammatau}
    \Gamma t_{0} = 3.72\times10^{-3} \left(\frac{m_s}{\textrm{keV}}\right)^5 \sin^2 2\theta~.
\end{equation}

The additional radiative decay channel $\nu_s\rightarrow \nu_a \gamma$ has a branching ratio of $\sim 1\%$. Its
a decay rate is~\cite{Pal:1981rm}
\begin{equation}
\label{eq:decaygamma}
    \Gamma _{\nu \gamma} = \frac{9 \alpha G_F^2}{2048\pi^4} m_s^5 \sin^2 2\theta \simeq 6.8 \times 10^{-33}~\textrm{s}~\Big(\dfrac{\sin^2 2\theta}{10^{-10}}\Big)\Big(\dfrac{m_s}{1~\textrm{keV}}\Big)^5~,
\end{equation}
where $\alpha \simeq 1/137$ is the fine structure constant. This decay channel could be the source of the putative $3.5$ keV X-ray line~\cite{Bulbul:2014sua,Boyarsky:2014jta}, and is also used to significantly constrain $\nu_s$ mixing angle for masses $m_s \gtrsim 1$ keV through astrophysical X-ray data as reviewed in Ref.~\cite{Abazajian:2017tcc}. 

For masses above the positron-electron pair production threshold $m_s > 2 m_e$, pair production $\nu_s \rightarrow \nu_a e^+ e^-$ is about $1/3$ of the $\Gamma_{3\nu}$~\cite{Picciotto:2004rp,Fuller:2011qy}. The total decay rate then becomes  about $4/3$ of the rate in Eq.~\eqref{eq:decaygamma}.

When $m_s$ is larger than the pion mass there are additional hadronic decays channels $\nu_s \rightarrow \pi^0 \nu_a$ as well as $\nu_s \rightarrow \pi^{\pm} e^{\mp}$. For $m_s > 2 m_{\mu} \simeq 211.4$~MeV, where $m_{\mu} = 105.7$~MeV is the muon mass, the decay modes $\nu_s \rightarrow \nu_a \mu^+ \mu^-$ become possible with the same rate as the decay into electron-positron pairs. For even larger $\nu_s$ masses, there are additional decay modes, such as $\nu_s \rightarrow \pi^{\pm} \mu^{\mp}$.

\section{The 511 keV INTEGRAL signal}
\label{appsec:511kev}

We add here a comment about the persistent 511~keV X-ray signal~\cite{Leventhal:1978,Johnson:1972} that has been observed with high precision by the SPI spectrometer aboard the INTEGRAL satellite~\cite{Knodlseder:2005yq,Siegert:2016ijv,Calore:2022pks} coming  from Galactic central region. A variety of explanations have been proposed for this signal (see e.g.~\cite{Boehm:2003bt,Beacom:2005qv,Wang:2005cqa,Totani:2006zx,Finkbeiner:2007kk,Bandyopadhyay:2008ts,Prantzos:2010wi,Wilkinson:2016gsy,Fuller:2018ttb,Fuller:2017uyd,Cappiello:2023qwl}). One could ask if the decays $\nu_s \rightarrow \nu_a e^+e^-$, possible for $m_s > 2 m_e$, could contribute to this signal through positron annihilation resulting in 511~keV radiation. The answer is that they could contribute only at the percent-level, given present upper limits on $\nu_s \rightarrow \nu_a \gamma$ decays~\cite{Calore:2022pks}  with the sterile-acting mixing orders of magnitude smaller, and DM fraction in $\nu_s$s larger, than if $\nu_s$s would be produced by the DW mechanism.

\bibliography{pbhneutrino}
\addcontentsline{toc}{section}{Bibliography}
\bibliographystyle{JHEP}
\end{document}